\documentclass[aps,prb,twocolumn,amsmath,nofootinbib,amssymb,superscriptaddress]{revtex4-1}
\usepackage{natbib}
\usepackage{graphicx}
\usepackage{color}
\usepackage[colorlinks, linkcolor=blue]{hyperref}
\usepackage{amssymb}
\usepackage{gensymb}
\usepackage{float}
\usepackage{amsmath}
\usepackage{tabularx,graphicx}
\usepackage{epstopdf}
\usepackage{latexsym}
\usepackage{color, colortbl}
\usepackage{psfrag}
\usepackage{bbm}
\usepackage{bm}
\usepackage{titlesec}
\usepackage{dsfont}
\usepackage{feynmp}
\usepackage{slashed}
\usepackage{multirow}
\usepackage{appendix}
\usepackage{ragged2e}
\usepackage{bbold}

\def \be{\begin{equation}}
\def \ee{\end{equation}}
\def \ba{\begin{array}}
\def \ea{\end{array}}
\def \bea{\begin{eqnarray}}
\def \eea{\end{eqnarray}}
\def \nn{\nonumber}
\def \ve {\varepsilon}

\def \W{{\Omega}}
\def \e{{\epsilon}}

\def \t{{\theta}}
\def \b{{\beta}}

\def \D{{\Delta}}
\def \d{{\delta}}

\def \s{{\sigma}}
\def \f{{\varphi}}

\def \e{{\epsilon}}
\def \ve{{\varepsilon}}

\def \ba{\begin{align*}}
\def \ea{\end{align*}}

\newcounter{indice}

\def \mrm{\mathrm}

\def \bs{\boldsymbol}
\def \mc{\mathcal}

\date{April 2021}

\begin{document}
\title{The Orbitally Selective Mott Phase in Electron Doped Twisted TMDs:\\ A Possible Realization of the Kondo Lattice Model}
\author{Amir Dalal}
\author{Jonathan Ruhman}
\affiliation{Department of Physics, Bar-Ilan University, 52900, Ramat Gan, Israel}
\affiliation{Center for Quantum Entanglement Science and Technology, Bar-Ilan University, 52900, Ramat Gan Israel}
\begin{abstract}
Moir\'e super-potentials in two-dimensional materials allow unprecedented control of the ratio between kinetic and interaction energy. 
By this, they pave the way to study a wide variety of strongly correlated physics under a new light. 
In particular, the transition metal dichalcogenides (TMDs) are promising candidate ``quantum simulators" of the Hubbard model on a triangular lattice. Indeed, Mott and generalized Wigner crystals have been observed in such devices. Here we theoretically propose to  extend this model into the multi-orbital regime by focusing on electron doped systems at filling higher than 2. As opposed to hole bands, the electronic bands in TMD materials include two, nearly degenerate  species, which can be viewed as two orbitals with different effective mass and binding energy. Using  realistic band-structure parameters and a slave-rotor mean-field theory, we find that an orbitally selective Mott (OSM) phase can be stabilized over a wide range of fillings, where one band is locked in a commensurate Mott state, while the other remains itinerant with variable density. This scenario thus, realizes the basic ingredients in the Kondo lattice model: A periodic lattice of localized magnetic moments interacting with metallic states. We also discuss  possible experimental signatures of the OSM state.
\end{abstract}
\maketitle

\section{Introduction}
Experiments in van der Waals materials have convincingly demonstrated the power of moir\'e super lattices as a tool to tune the strength of electronic correlations. Following the theoretical prediction~\cite{bistritzer2011moire} a wide variety of strongly correlated phenomena was experimentally observed~\cite{cao2018unconventional,yankowitz2019tuning,lu2019superconductors,balents2020superconductivity,cao2018correlated,chen2019evidence,lu2019superconductors,jiang2019charge,kerelsky2019maximized,cao2020nematicity,zondiner2020cascade,wong2020cascade,lu2019superconductors,chen2020tunable,serlin2020intrinsic,zondiner2020cascade,wong2020cascade}. The Dirac dispersion, characterizing the unperturbed electronic states in graphene, leads to topologically non-trivial flat bands \cite{po2018origin,song2019all} with large Wannier orbitals~\cite{po2018origin,kang2018symmetry,yuan2018model}, from which these correlated states emerge.

\begin{figure} [!h]
  \centering
  \includegraphics[width=0.475\textwidth]{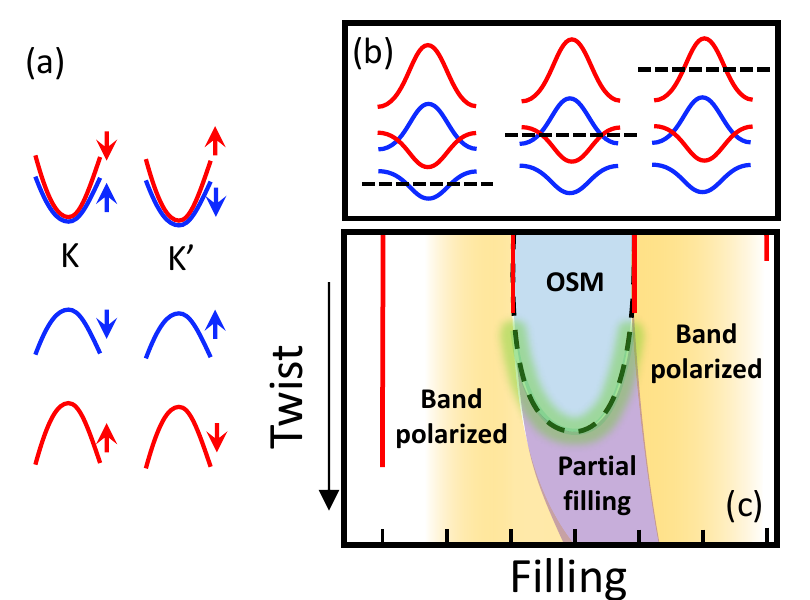}
  \caption{   (a) Schematic band structure of a single-layer TMD near the $K$ and $K'$ points. While the hole Bloch bands have a large SOC splitting, the electron bands are nearly degenerate (red and blue indicate different Bloch bands, which we refer to as ``species"). (b) With a moir\'e potential the electron bands form multiple flat minibands that can overlap in energy space and be simultaneously at partial filling. (c) Schematic phase diagram arising from our slave-rotor mean-field analysis. A charge localized state of one species can coexist with a Fermi liquid state of the other, which is known as the orbitally selective Mott (OSM) state. Inside the region marked by the dashed black line the essential ingredients of a Kondo lattice model are realized. The red lines indicate correlated insulating states.   \label{fig:schem_PD}}
\end{figure}

In contrast, semiconducting transition metal dichalcogenides (TMDs) subject to moir\'e potentials are expected to have a simpler microscopic picture. The low energy physics is captured by a Hubbard model on a triangular lattice~\cite{wu2018hubbard,zhang2019moir}. Mott insulators and generalized Wigner crystals have been experimentally observed~\cite{wang2020correlated,regan2020mott,xu2020correlated,xu2020abundance,li2021continuous}, as well as possible indications of superconductivity~\cite{wang2020correlated}. The relative simplicity of their microscopic starting point makes the TMD moir\'e devices prime candidates for condensed matter ``quantum simulators" of the Hubbard model. 

A canonical model that is both of great fundamental interest to quantum condensed matter, and has not yet been realized in moir\'e devices, is the Kondo lattice model~\cite{doniach1977kondo}. Its main ingredients are a lattice of localized moments coupled to a Fermi liquid of itinerant electrons. The main coupling between these two degrees of freedom is spin-exchange. 
The case where the strongest exchange mechanism is antiferromagnetic is understood to be the minimal model that captures the low-energy physics of many rare-earth compounds, known as {\it heavy-fermion materials}~\cite{steglich1979superconductivity,stewart1984heavy}. When the dominant exchange is the Hund's coupling between the local and itinerant orbitals, the coupling is ferromagnetic. Such a scenario was discussed in the context of the {\it orbitally selective Mott} (OSM) phase~\cite{nakatsuji2000quasi,anisimov2002orbital}.

Materials that host coexisting itinerant and localized states, exhibit a plethora of exotic phases such as heavy-fermi liquids, metallic magnets, high-$T_c$ superconductors and non-Fermi liquids~\cite{lohneysen1998heavy,von1996non,von1998magnetic,nakatsuji2000quasi,schroder2000onset,anisimov2002orbital,senthil2003fractionalized,Stewart2001non,custers2003break,custers2003break,senthil2004weak,coleman2007heavy,gegenwart2008quantum,vojta2010orbital}.
However, what makes them especially interesting is the existence of quantum phase transitions, where the lattice of local moments melts into a metallic state~\cite{schroder2000onset,lohneysen1998heavy,custers2003break,yuan2003observation,gegenwart2008quantum,aoki2019review,jiao2020chiral,li2021continuous}. Such a transition is not captured by the Ginzburg-Landau paradigm because it must include a whole Fermi surface that emerges at the quantum critical point~\cite{senthil2003fractionalized,senthil2004weak,vojta2010orbital}.
What controls the different ground states, and the nature of the quantum critical point separating them, is still debated. However, the comparison between theory and experiment becomes challenging  due to the complex structure of the materials which realize this physics. 
For this reason, a controlled experimental realization of such a minimal model is highly desirable.

\begin{figure}
  \centering
  \includegraphics[width=0.45\textwidth]{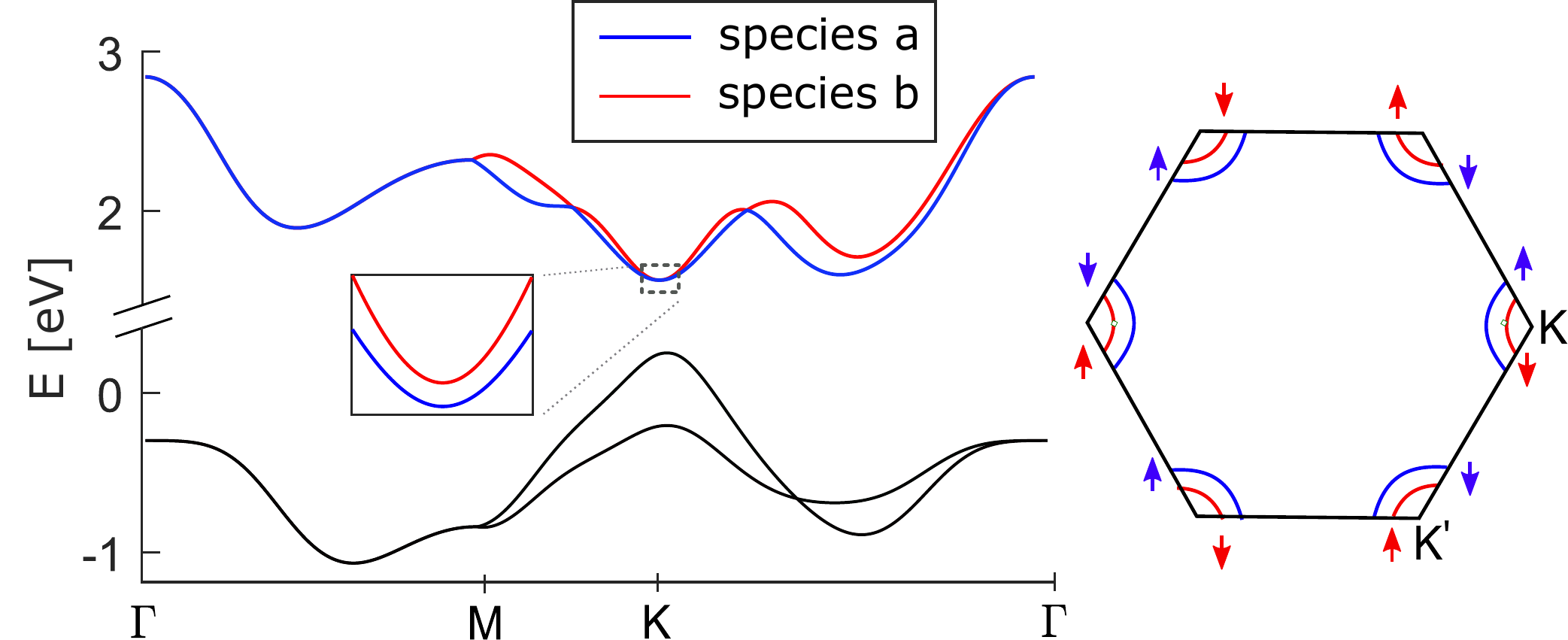}
  \caption{Left: Typical dispersion of the two lowest Bloch bands of the conduction band in a bare single-layer TMD obtained from the tight-binding model~\cite{liu2013three-band}.  Near the $K$ and $K'$ points the bands are approximately parabolic and assume a small splitting due to spin orbit coupling in the second order (see inset). Right: Upon lightly doping the system two Fermi pockets of opposite spin orientation form around each high symmetry point, corresponding to the two Bloch bands.     \label{fig:bands}}
\end{figure}

\begin{table}
\caption{\label{tab:table1}
The effective mass and spin orbit splitting near the conduction band minima of the high-symmetry points $K$ and $K'$ for different TMD single layers (taken from Ref.~\onlinecite{liu2013three-band}). Here $\D = \D_b-\D_a$ is the spin-orbit splitting between the bands.}
\begin{ruledtabular}
\begin{tabular}{cccccccc}
 &$m_{a}/m_{e}$ &$m_{b}/m_{e}$ &$\Delta$ (meV)\\
\hline
MoS$_{2}$& 0.45 & 0.51 & 3&\\ 
MoSe$_{2}$& 0.51 & 0.61 & 21 \\
WS$_{2}$& 0.40 & 0.30 & 29 \\
WSe$_{2}$& 0.44 & 0.31 & 36 \\
\end{tabular}
\end{ruledtabular}
\\
\end{table}

\begin{figure}[t]
  \centering
  \includegraphics[width=0.475\textwidth]{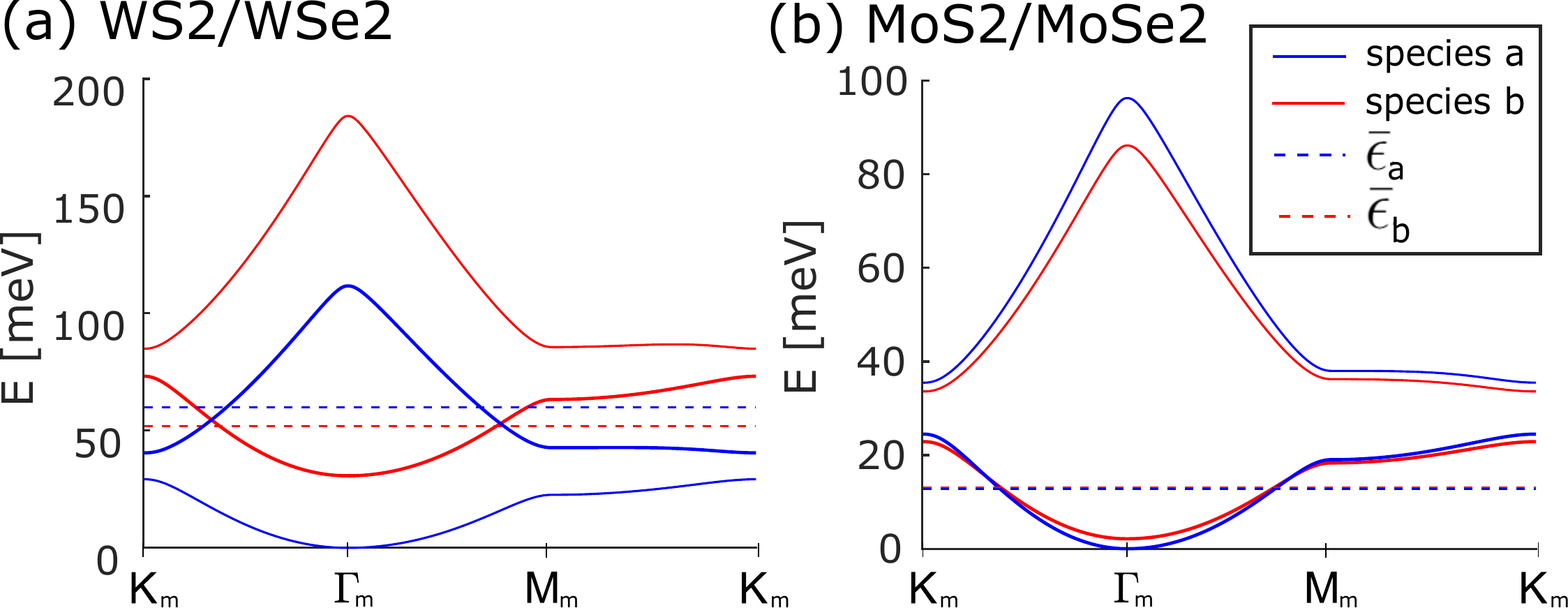}
  \caption{ The two lowest minibands of the two species near the conduction band minima resulting from a moir\'e potential of angle $\theta_M =3^\circ$ and depth $V_0 = 15$ meV. The dashed lines indicate the average band energies, $\bar \e_\tau$ (see text). In this paper we focus on the lowest pair of such minibands that {\it overlap} in energy space.  (a) For WS$_2$/WSe$_2$ this pair includes the lowest miniband of species $b$ and the first excited band of species $a$. (b) For MoS$_2$/MoSe$_2$ the SOC is very weak and the miniband structure is nearly degenerate for all minibands. Therefore, for this material we consider the lowest miniband for both Bloch bands. 
   \label{fig:minibands}}
\end{figure}

 In this paper we explore the conditions under which, TMDs subject to a moir\'e potential can host a  state of coexisting itinerant and localized electrons known as the orbitally selective Mott state (OSM).\cite{anisimov2002orbital,biermann2005non,vojta2010orbital}
We first argue  that the mini-band structure of {\it electron doped} TMD moir\'e devices can potentially host multiple flat bands, which can be simultaneously at a state of partial filling. 
This is mainly because of the relatively small spin-orbit splitting of the bare conduction bands around the $K$ and $K'$ points~\cite{liu2013three-band} [see Fig.~\ref{fig:schem_PD}.(a)]. 
We consider the situation where such minibands are induced in one layer by another ``inactive" layer in a heterogeneous structure~\cite{wu2018hubbard}. For example, two prototypical bilayers we consider are WS$_2$/WSe$_2$, where the effects of spin-orbit splitting on the conduction bands are small but noticeable, and MoS$_2$/MoSe$_2$, where the splitting is negligible. In both cases the sulfur based compounds are where the electronic states reside, and the selenium based layers take the role of the ``inactive" layer that induces the moir\'e potential. 
Using a slave-rotor mean-field approximation~\cite{Florens2004slave,zaho2007self,chen2020competition} within a simplified on-site interaction Hubbard model, we identify the emergence of the OSM state at fillings surrounding $n=4$ or $n=2$ (depending on the strength of spin-orbit coupling). In this phase one species is in a Mott state, while the other species is partially filling one of its minibands [see Fig.~\ref{fig:schem_PD}.(b)\&(c)].

\section{ Model Hamiltonian}
The two lowest Bloch bands above the band gap, which we denote here as species $\tau = a,b$, have nearly degenerate band minima in the vicinity of each valley, $K$ and $K'$~\cite{liu2013three-band}(see Fig.~\ref{fig:bands}). Due to spin-orbit coupling they are split and assume different effective masses.
As mentioned above, this splitting is significantly smaller compared with the equivalent splitting in the valence band.
Nonetheless, the spin projection along $z$ remains a good quantum number up to second order in perturbation theory.

As usual we obtain an additional valley degree of freedom by expanding the momentum around $K$ and $K'$. Note however, that spin-orbit coupling slaves spin to valley within a given Bloch band. We therefore denote the additional degree of freedom by its spin as follows: 
For the lower Bloch band $\tau = a$, the state $\s = \uparrow$ and $\s = \downarrow$ corresponds to a valley $K$ and $K'$, respectively.  On the other hand, for the higher Bloch band $\tau = b$, the state $\s = \uparrow$ and $\s = \downarrow$ corresponds to a valley $K'$ and $K$, respectively.

The resulting Hamiltonian (up to quadratic order in deviation from the high-symmetry points), is given by  
\begin{eqnarray}\label{eq:H0}
\mc {H}_0 = \sum_{\bs k\tau\s}\left(\Delta_{\tau}+\frac{k^2}{2m_\tau}\right){c}^{\dagger}_{\bs k\tau\s}{c}_{\bs k\tau\s}\,.
\end{eqnarray}
Where $\D_\tau$ and $m_\tau$ are the species dependent band minimum and mass, respectively. The values are listed in Table~\ref{tab:table1}.

In principle, the Fermi surfaces surrounding the $K$ and $K'$ points (in terms of the momentum relative to these points), are non-degenerate except for six high symmetry lines. 
However, as a result of the parabolic band approximation, used in Eq.~\eqref{eq:H0}, the Fermi surfaces are spherically symmetric and are thus doubly degenerate everywhere (the degeneracy corresponds to the spin index $\s$). This reflects an emergent SU(2) symmetry of each species~\cite{wu2018hubbard}.\\

In this paper we consider two prototypical TMD bilayers WS$_2$/WSe$_2$ and MoS$_2$/MoSe$_2$. In both cases the band alignment properties are such, that the charge carriers reside on the sulfur based side of the bilayer upon electronic doping. Thus, the selenium based layers are inactive and only induce the moir\'e potential. In the case of MoS$_2$/MoSe$_2$ spin-orbit coupling is very weak and consequently the masses $m_\tau$ and band minimum points $\D_\tau$ in Eq.~\eqref{eq:H0} are almost identical. 
In the case of  WS$_2$/WSe$_2$, the effects of spin-orbit coupling are more noticeable, such that $m_b/m_a \approx 0.75$ and $\D_b - \D_a = 30$ meV.  

We now turn to consider the influence of a moir\'e potential on the band structure close to the bottom of the conduction bands $
\tau = a,b$. We follow Refs.~\onlinecite{wu2018hubbard,zhang2019moir}. The induced potential is given by 
\begin{align}\label{eq:H_M}
\mc H_{M}=\sum_{j\tau\bs k \s}V_{0}(\pmb{G}_j) {c}^\dagger_{\bs k+\bs G_j\tau\s}{c}_{\bs k \tau\s}\,,
\end{align}
where  $\pmb{{G}}_{j}=\hat{R}(j\frac{\pi}{3})(4\pi/\sqrt{3}a_{M}\hat{x})$, $j=0,\ldots,5$ are the six shortest reciprocal lattice vectors of the moir\'e super lattice. $a_{M}=a/\t_M$ is the moir\'e lattice constant and $\t_M = \sqrt{\d^2 + \t^2}$ is the effective twist angle. Here $\d$ is the lattice mismatch and $\t$ accounts for any additional twist.  We take the strength of the potential to be $V_0 = 15$ meV for both bilayers. 

The parabolic Hamiltonian Eq.~\eqref{eq:H0} together with the moir\'e potential Eq.~\eqref{eq:H_M} are diagonalized using a nearly-free electron approximation truncated at the level of 19 bands (3rd nearest neighbour in reciprocal space). 

In Fig.~\ref{fig:minibands} we plot the two lowest minibands of each species using realistic parameters for the bilayers. In panel (a) we show that for the strongly spin-orbit coupled bilayer, WS$_2$/WSe$_2$, the lowest miniband of species $b$ overlaps with the first remote miniband of species $a$. On the other hand, in panel (b) we show that for the weakly spin-orbit coupled bilayer, MoS$_2$/MoSe$_2$, the minibands of the two species are almost identical. In this case the two lowest minibands (and the two first excited bands) overlap in energy. 

We will be interested in the physics arising from partially filling two different minibands simultaneously. Therefore, from here on we will focus exclusively on the {\it lowest pair of mninbands that overlap in energy space} corresponding to the two Bloch bands $\tau = a,\,b$. 
The miniband Hamiltonian then assumes the form 
\begin{eqnarray}\label{eq:H_b}
\mc H_{b}=\sum_{k\tau\s}\xi_{\bs k\tau}\psi_{ k \tau\s}^\dagger{\psi_{k\tau\s}}
\end{eqnarray}
However, we still define the density in units of {\it total filling} starting from the bottom of the conduction band. Consequently, for WS$_2$/WSe$_2$ [Fig.~\ref{fig:bands}(a)] the relevant range of filling is $n\in[2,6]$, where the lowest miniband of species $a$ is already {\it completely filled} and contributes a background charge of 2. This situation is also depicted in the center of panel (b) in Fig.~\ref{fig:schem_PD}. On the other hand for MoS$_2$/MoSe$_2$ the two lowest minibands of each species overlap and therefore we focus on the range of filling $n\in[0,4]$.

The second ingredient in our model is the interaction. We consider an on-site repulsion of the form
\begin{align}\label{eq:H_I}
\mc H _I = {U\over 2}\sum_{i}(\eta\,\delta n_{ia}+\delta n_{ib})^2\
\end{align}
where  $\delta n_{\tau} = \sum_{\s}\psi_{\tau\s}^\dag \psi_{\tau\s} -1$ is the density operator in particle-hole symmetric form.\footnote{This shift can be absorbed into $\bar\e_\tau$.} $\eta$ is a phenomenological parameter, which accounts for the possible difference in the Wannier-orbital spread of the species. When the lowest miniband of species $b$ overlaps with the first remote miniband of $a$ (Fig.~\ref{fig:minibands}. a), the spread of the Wannier orbital of the latter is expected to be wider than that of the former. Consequently, electrons in miniband $a$ will have a weaker Coulomb repulsion, corresponding to $\eta<1$.  
On the other hand, when the overlapping minibands are both the first flat band (Fig.~\ref{fig:minibands}. b)  the interactions are expected to be roughly equal and $\eta = 1$.  We consider a constant interaction $U$ of moderate strength, which corresponds to the estimate of Ref.~\cite{wu2018hubbard} with a large dielectric constant $\kappa \approx 5$~\cite{laturia2018dielectric}.~\footnote{We neglect the angle dependence of $U$.}

\begin{figure*}
  \centering
  \includegraphics[width=1\textwidth]{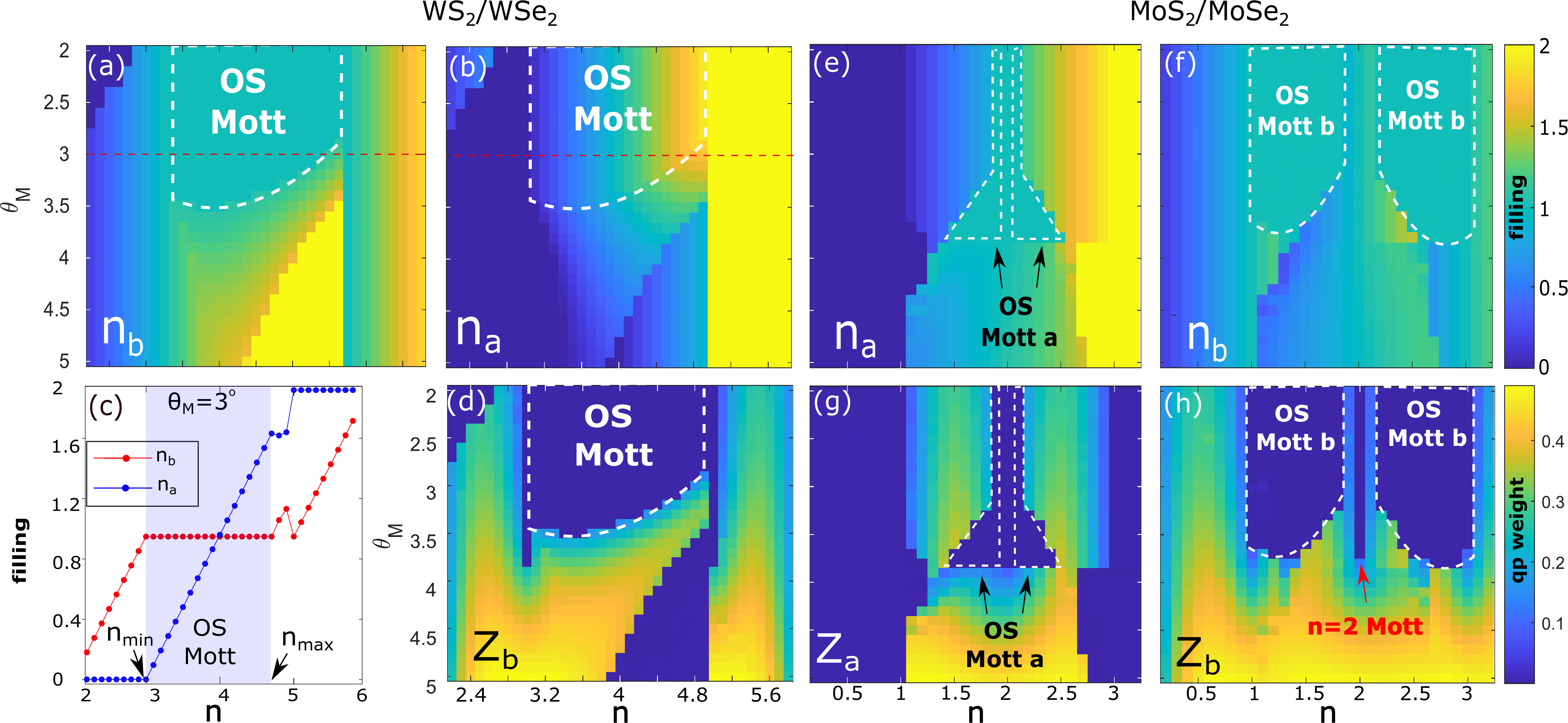}
  \caption{Results of the slave-rotor mean-field theory. (a)-(d) Results for WS$_2$/WSe$_2$ at filling higher than 2 using $U=60$ meV and $\eta = 1/2$. (a) Colormap of the density of band $b$ in the space of filling $n$ and twist angle $\t_M$. (b) The corresponding density map of band $a$. (d) Colormap of the quasiparticle weight of species $b$. (c) The densities of species $a$ and $b$ at $\t_M=3^\circ$ vs. total density.
  (e)-(h) Results for MoS$_2$/MoSe$_2$ using $U = 40$ meV and $\eta=1$. (e) \& (f) The density of species $a$ \& $b$, respectively. (g) \& (h) are maps of the quasiparticle weights of species $a$ \& $b$, respectively. In all cases we observe finite regions, where the quasiparticle weight of a certain band falls to zero concomitant with odd integer filling. In these regimes the other species is in a compressible Fermi liquid state and supports non-integer filling. These regimes thus, realize the OSM and are marked by white dashed lines.   \label{fig:phase_diagram}}
\end{figure*}

\section{Slave Rotor Mean-Field analysis }
We now turn to study the ground state of the Hamiltonian Eqs.~(\ref{eq:H_b},\ref{eq:H_I}). In particular, we are interested to understand whether a Mott state of one of the species can be stabilized over a finite density range, where the other band remains metallic. 
To this end, we employ the slave-rotor mean field theory~\cite{Florens2004slave,zaho2007self,chen2020competition}.  
It consists of decomposing the field operators 
into bosonic rotors multiplied by neutral spinon operators 
$
\psi_{i\tau\s}=e^{-i\theta_{i\tau}}\,f_{i\tau\s} 
$.
The respective density operator of each species, which are conjugates of the phases above, are then represented by angular momentum operators 
$
\hat{L}_{i\tau}=-i{\partial}/{\partial\theta_{i\tau}}
$,
subject to the {\it local} constraint
$
\hat{L}_{i\tau}= f^\dagger_{i\tau}f_{i\tau}-1 
$,
where the sum over spin is implicit.

The above decomposition allows for a mean-field treatment of the Mott transition~\cite{Florens2004slave}. The corresponding ``order parameter" is the quasiparticle weight $Z_\tau = |\langle e^{i\theta_\tau}\rangle |^2$. When the rotor fields are pinned $Z_\tau\ne 0$, resulting in a finite overlap between the quasiparticle  and bare-electron operators.
Moreover, the uncertainty principle implies the conjugate charge operator $\hat L_\tau$ experiences large fluctuations. Thus, we can identify this phase with a Fermi liquid. 
On the other hand, when the charge operators $\hat L_\tau$ are pinned, which corresponds to small charge fluctuations,  the conjugate phases are strongly fluctuating and $Z_\tau=0$. This phase is thus associated with the Mott state.

Before applying the slave-rotor decomposition however, it is essential to decompose the miniband dispersion, $\xi_{\bs k \tau}$, Eq.~\eqref{eq:H_b} into two terms
\be\label{eq:disp}
\xi_{\bs k\tau} = \bar \e_\tau +  \e_{\bs k \tau}\,,  
\ee
where $\bar \e_\tau$ is the average energy of the miniband ($\bar \e_\tau = \sum_{\bs k\in MBZ}\xi_{\bs k,\tau}$), and the remainder, $\epsilon_{k\tau}$, is the kinetic  part of the dispersion, which averages to zero. 
$\bar \e_\tau$ can be interpreted as the effective binding energies of electrons to  the respective minibands (dashed lines in Fig.~\ref{fig:minibands}).
The importance of this decomposition, is to separate these local energy shifts from the dispersion because they should not be renormalized by the quasiparticle weights $Z_\tau$. Indeed, in the slave-rotor theory the quasiparticle weight only renormalizes the band width but does not shift the average energy of the band~\cite{Florens2004slave}.


Performing  the slave-rotor decomposition to both species we obtain the Hamiltonian
\begin{align}\label{eq:H_SR}
\mc H_{SR}=&- \sum_{ij\tau\s}\left(t_\tau^{ij} e^{i(\theta_{i\tau}-\theta_{j\tau})}+\d_{ij}\bar \e_\tau \right)f^\dag_{i\tau\s}f_{j\tau\s}
\\
&+{U\over 2}\sum_{i} \left( \hat L_{ia} + \eta \hat L_{ib} \right)^2
\nn
\end{align}
where $t^{ij}_\tau$ are the set of tight-binding parameters that reproduce the dispersive part $\e_k$ in Eq.~\eqref{eq:H_M} when transformed to reciprocal space.

To asses the ground state of the Hamiltonian Eq.~\eqref{eq:H_SR} we employ the variational method, as opposed to Ref.~\onlinecite{Florens2004slave}, where the self-consistent mean-field approach was used. Namely, we minimize the expectation value of Eq.~\eqref{eq:H_SR} with respect to the variational wavefunction denoted by $|\W_V\rangle = |K_a,h_a\rangle\otimes |K_b,h_b\rangle\otimes|\mu_a\rangle\otimes |\mu_b\rangle$. This variational state is a product of the ground-states of the rotor Hamiltonians
\be\label{eq:H_theta}
H_{\theta}^
\tau=\frac{1}{2}{\hat L_
\tau}^2+h_\tau\hat{L}_{\tau}+K_{\tau}\cos\theta_{\tau}
\ee
and two Slater-determinant  states ("Fermi sea" states) of the spinons, where the density is controlled by the chemical potentials $\mu_a$ and $\mu_b$.

We must determine  six variational parameters with three constraints $\langle \hat L_\tau \rangle = \langle f^\dag_{i\tau}f_{i\tau} \rangle-1$ and $\sum_{\tau}\langle  f^\dag_{i\tau} f_{i\tau}\rangle = n$. 
The parameter $K_
\tau$ controls whether species $\tau$ is metallic or localized. When the minimal energy solution is obtained with $K_\tau\ne 0$ the rotors are pinned and we get a finite quasiparticle weight $Z_\tau \ne 0$ corresponding to the metallic state. On the other hand, for $K_\tau = 0$ the rotors are in eigenstates of the angular momentum operator where the average of $e^{i\t_\tau}$ vanishes, corresponding to the Mott state ($Z_\tau = 0$). Additionally, there is a freedom to redistribute charge between the two bands, which is controlled by the difference in the chemical potentials, $\mu_a-\mu_b$. Finally, the constraints are fulfilled using the three Lagrange multipliers $h_\tau$ and the sum $\mu_a+\mu_b$. 
Note that in the case of WS$_2$/WSe$_2$, where the overlapping bands include one the first excited band of species $a$, which is more disprersive compared to the lowest band of species $b$, we apply the slave-rotor decomposition only to band $b$. 
For more details on the slave-rotor analysis we perform and the minimization procedure see Appendix~\ref{app:slave-rotor}.

In Fig.~\ref{fig:phase_diagram} we plot the phase diagrams resulting from the variational minimization. Panels (a)-(d) correspond to the WS$_2$/WSe$_2$ bilayer using $U = 60$ meV and $\eta = 1/2$. Panel (a) and (b) are maps of the filling of species $b$ and $a$, respectively, in the space of total filling $n$ and twist angle $\t_M$. In this case we recall that there is another completely filled miniband below the relevant pair of overlapping minibands (see Fig.~\ref{fig:minibands}.a), therefore the total filling is given by 
$n = 2+n_a+n_b\,.$
We turn our focus to the region inside the white dashed line, where the filling of species $b$ is locked to unity, while the filling of species $a$ varies continuously. \footnote{One should note the regions where the densities of the different species are locked to the value  $n_\tau = 2$. These regions correspond to a band insulator in the relevant species, where the corresponding band is completely filled with two particles per moir\'e unit cell.}
In the same region, we find that the quasiparticle weight $Z_b$ vanishes [panel (d)]. Thus,  this region is identified as the OSM phase, where a lattice of localized magnetic moments coexists with itinerant electrons. 
Panel (c) shows the  filling of each band for a specific twist angle $\t_M = 3^\circ$ showing that the density of the itinerant band can be tuned over a large range inside the OSM phase.

In panels (e)-(h) we plot the results for MoS$_2$/MoSe$_2$ using $U = 40$ meV and $\eta = 1$. Panels (e) and (f)  are maps of the filling of species $a$ and $b$, respectively. Note that in this case the total density is simply $n = n_a+n_b$ Panels (g) and (h) are the quasiparticle weights of bands $a$ and $b$, respectively. Here we identify two OSM phases, one where band $a$ is locked in a Mott state (OSM$_a$) and one where band $b$ is locked in the Mott state (OSM$_b$). Additionally, at $n = 2$ there is a Mott state of both species, which is expected to have an approximate SU(4) symmetry~\cite{keselman2020dimer,keselman2020emergent,wu2019topological,zhang2021su4}.

For both materials, the OSM state assumes a large portion of the phase space (in Appendix \ref{app:slave-rotor}  we show that this scenario is relevant to other TMD materials).
For small twist angles, the density of itinerant electrons can be tuned between  completely empty and completely filled states, which potentially allows to tune the strength and sign of the RKKY interaction between local moments, which is mediated by the itinerant miniband~\cite{fischer1975magnetic}. At larger twist angles ($\t_M \gtrsim 4^\circ$) the lattice of localized electrons melts into a Fermi liquid. Such a transition is characterized by the emergence of Fermi surface, which is not captured by the Ginzburg-Landau paradigm and is therefore of special interest~\cite{senthil2004weak,senthil2003fractionalized,vojta2010orbital}.\footnote{ We note that this transition line will be pushed to larger $\t_M$ upon increase of the interaction parameter $U$.}  

We have also tested the stability of the OSM phase to variations in the parameters $\D \bar \e = \bar \e_a-\bar \e_b$ and $\eta$ numerically. In Appendix~\ref{app:stab} we show that the range of filling where the OSM phase occurs is large for a wide range of $\D \bar \e$ and $\eta$. We also roughly estimate this range analytically and find that it is expected to be large in the parameter regime $|\D \bar \e| < U(1-\eta)^2/2-T_{it}$. Here $T_{it}<0$ is the the kinetic energy gain of filling the Fermi sea of the itinerant band minus the interaction energy associated with onsite fluctuations of charge. This analysis shows that the existence of a wide OSM phase is robust to the parameters of our model. 

\begin{figure}[!b]
  \centering
  \includegraphics[width=0.35\textwidth]{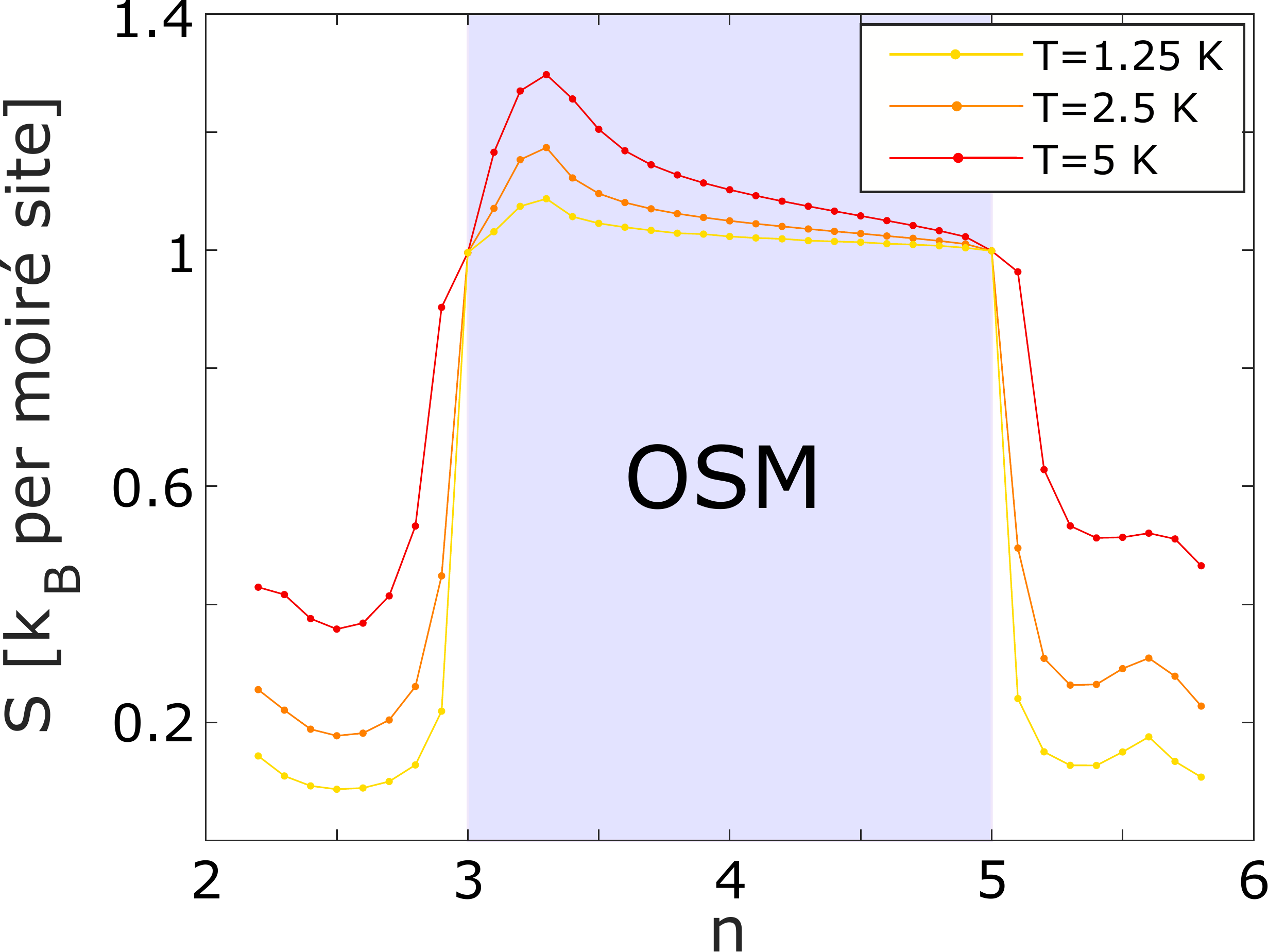}
  \caption{The entropy per site in units of $k_B \log 2$ as a function of filling at angle $\t_M = 2.5 ^\circ$ for three different temperatures. Here the Mott state is assumed to contribute one bit per site.  \label{fig:entropy}}
\end{figure}

\noindent
\section{ Experimental consequences of the OSM state}
We turn to discus experimental consequences of the OSM phase.  We first discuss the enlarged entropy associated with the formation of local moments. When the moments are free they contribute one $k_B$  per lattice site. If a magnetic ordering is present, the local-moment contribution will be significant above the ordering temperature. 
Additionally, a distinct feature of this contribution will be a strong dependence on magnetic field. 
Indeed, the authors of Refs.~~\onlinecite{rozen2020entropic,saito2020isospin} have recently measured such an enlarged entropy in TBG, where they attributed it to local moments coexisting with metallic states.  Similarly, in the regime where both phases are metallic, but close to the OSM regime we may expect a Pomeranchuk effect upon heating~\cite{rozen2020entropic,saito2020isospin}.

To estimate the change in entropy across the OSM transition we assume the local moments contribute their maximal entropy, while the metallic states contribute 
$
{s_M} = -{2\over \W} \sum_{\tau,\bs k \in BZ} \left[ N_{k\tau}\log N_{k\tau}+(1-N_{k\tau})\log(1-N_{k\tau}) \right]
$, where $N_{k\tau}$ are the momentum space Fermi-Dirac distribution functions, which include the effects of the quasiparticle weight $Z_\tau$.
In Fig.~\ref{fig:entropy} we plot the entropy per moir\'e lattice site as a function of density for the WS$_2$/WSe$_2$ bilayer at $\t_M = 2.5^\circ$ for three different temperatures. The distinct signature is a large jump at the boundaries of the OSM state, where we also observe an enhanced specific heat manifested in the strong dependence of $S$ with $T$.  

Another suitable probe for the OSM state is magneto-transport~\cite{vojta2010orbital}, especially given that we predict this state at a relatively high density range $n\sim 2-4$, where the effects of disorder are less prominent as compared with the filling range of the lowest miniband.
Inside the OSM phase the Hall number, which is seen both in the slope of the classical Hall resistivity and in the period of quantum oscillations, will correspond to a ``small" Fermi surface (of volume $n_b = n-1$). 
At the phase transition point [green hue in Fig.~\ref{fig:schem_PD}.(c)] the local moments melt into a metallic state, manifested in a Lifshitz transition, where we can distinguish two scenarios. When the dominant exchange interaction between the two species is antiferromagnetic we expect a heavy-fermi liquid state to emerge between the fully metallic phase and the magnetic metal.
In this case the Hall number changes from the ``small" volume $n-1$ to the ``large" volume $n$. 
The second scenario, is where the exchange interaction is dominated by ferromagnetic exchange (e.g. due to the orbital Hund's coupling). In this case, theory does not predict the emergence of a hybridization gap between the local and itinerant electrons. Instead, a new Fermi surface emerges at the transition point. Thus, we expect the appearance of beating in quantum oscillations and non-linearity of the classical Hall effect (see for example Ref.~\onlinecite{joshua2012universal}). Thus, magneto-transport measurements across the melting transition can also distinguish the nature of magnetic exchange mechanism.

\section{ Summary }
We proposed that electron doped TMDs subject to a moir\'e potential are prime candidates to realize the Kondo lattice model. The essential ingredient is the multiplicity of electron band minima close to the $K$ and $K'$ points, which allows for two moir\'e bands of different width to be simultaneously at partial filling. 
We used a simplified model with constant on-site Coulomb repulsion and a slave rotor mean-field theory to study the possible ground states of the system. We found a large phase space, where an orbitally selective Mott phase forms. Such a state is characterized by a Mott state of one species coexisting with a metallic state of the other. This opens a path to simulate the Kondo lattice model and possible exotic phase transitions in TMD mor\'e devices.

\noindent
{\it Note added --} Upon completion of this paper we came to learn about a related theoretical proposal regarding tri-layers of twisted graphene sheets~\cite{ramires2021emulating}.

\section{Acknowledgments}
We are grateful to Erez Berg, Debanjan Chowdhury, Rafael Fernandes, Efrat Shimshoni, 
Inti Sodemann and Arun Parameknti for helpful discussions.
This research was funded by the Israeli Science Foundation under grant number 994/19. JR acknowledges the support of the Alon fellowship awarded by the Israel higher education council. 

\appendix
\section{Continuum dispersion}\label{app:band_structure}
In this appendix we provide additional information about the computation of the continuum Hamiltonian.
We start with the tight-binding approximation for  single layer semiconducting TMDs of the trigonal prismatic structure (H) ~\cite{liu2013three-band}. This model consists of three orbitals $d_{z^2},d_{xy}$ and $d_{x^2-y^2}$, taking into account spin-orbit coupling and hopping up to the third nearest neighbours on the triangular lattice. 
 The conduction band consists of two Bloch bands denoted by $\tau = a,b$, which are plotted in Fig.~\ref{fig:bands} (colored red and blue, respectively) and will be referred to as ``species" henceforth.
Each such band has two parabolic minima near the $K$ and $K'$ corresponding to spin states $\s = \uparrow \downarrow$ (valley and spin are locked. However, it is important to note that the spin orientations near $K$ and $K'$ are opposite in the two Bloch bands). Up to quadratic order in deviations from the high-symmetry points we obtain the Hamiltonian
\begin{eqnarray}\label{eq:app:H0}
 \hat{H}_0 = \sum_{k\tau\s}{c}^{\dagger}_{k\tau\s}\left(\Delta_{\tau}+\frac{k^2}{2m_\tau}\right){c}_{k\tau\s}\,.
\end{eqnarray}
Here $\D_\tau$ and $m_\tau$ are species dependent band minimum  and mass, respectively. $\bs k$ is the lattice momentum relative to the high symmetry points, i.e. relative to $\bs K$ for $(a,\uparrow)$, $(b,\downarrow)$ and relative to $\bs K'$ for $(b,\uparrow)$, $(a,\downarrow)$. 
A crucial feature, which is unique to the conduction bands, is that the higher order spin-orbit splitting, $|\D_a - \D_b|$, is comparable to the expected moir\'e lattice depth and resulting miniband width (see Table.~\ref{tab:table1}).

We now turn to consider the effect of a moir\'e potential, which we assume is induced by a second layer. At small twist angles $\theta_M\ll \pi$ the superlattice constant is given by $a_{M}\approx{a}/{\t_M}$,
where $\t_M \equiv \sqrt{\delta^2+\theta^2}$, $\delta$ is the miss-match between the layers taken from Ref.~\onlinecite{zhang2019moir} and $\t$ accounts for any additional twist. In this limit we have  $a_{M}\gg a$, which justifies the use of a simple triangular periodic potential constructed out of the six smallest harmonics $\pmb{{G_{j}}}=\hat{R}(j\frac{\pi}{3})(4\pi/\sqrt{3}a_{M}\hat{x})$, $j=0,\ldots,5$:
\begin{eqnarray}
V_{M}(\pmb{r})=\sum_{i}V_{0}(\pmb{G_{i}})e^{-i\pmb{G_{i}}\cdot\pmb{r}}
\end{eqnarray}
The potential has three fold rotational symmetry which states that:
$V_0(\hat{R}(\frac{2}{3}\pi)\pmb{G_{i}})=V_0(\pmb{G_{i}})$  and
$V_0(-\pmb{G_{i}})=V^{*}_0(\pmb{G_{i}})$.
 To obtain the miniband structure of the lowest mini-bands we use a 19 band model without counting degeneracy of spin and species. For simplicity we take the moir\'e potential strength  to be uniform across platforms and given by $V_0 = 15$ meV~\cite{wu2018hubbard,zhang2019moir}.

In Fig. \ref{fig:minibands} we compare the two lowest minibands for the two species (species $a$ colored blue, and species $b$ colored red) for realistic parameters of two candidate materials.  As can be seen a feature of these miniband structures is the overlap of bands belonging to different species.
Note that the overlapping minibands are not necessarily the same numeral sub band. As shown in panel (a) for WS$_2$ the overlap is between the first excited band of one species and the lowest miniband of the other (the same is true for WSe$_2$ and MoSe$_2$). On the other hand for MoS$_2$ the overlap occurs between the lowest minibands of the two species (panel b).  Upon restriction to the two bands of interest (namely, those that are  overlapping) we obtain the dispersion in Eq.\eqref{eq:H_b}.

\section{Interactions}
In the paper we assume a contact interaction of the form
\begin{align}\label{eq:H_I1}
\mc H_I= {1\over 2}\sum_{i\tau\tau'}{U_{\tau \tau'}} \d n_{i\tau}\d n_{i\tau'}\,,
\end{align}
where $\d n_{i\tau} = \psi^\dag_{i\tau}\psi_{i\tau}-1$. Notice that we have written the interaction in a particle-hole symmetric manner, which can be absorbed into the parameters $\bar \e_\tau$ in Eq.\eqref{eq:disp}. 

The relative strength of the interaction parameters $U_{\tau \tau'}$ depend on the spread of the Wannier orbitals of the corresponding minibands~\cite{wu2018hubbard,zhang2019moir}. Namely, when both  minibands are the lowest sub-band of their corresponding species [as shown for MoS$_2$ in Fig.\ref{fig:minibands} (b)], the spread of the two Wannier  functions is approximately the same, and we expect $U_{aa}\simeq U_{ab}\simeq U_{bb}$. In this case the interaction \eqref{eq:H_I1} is proportional to the square of total density.

On the other hand, when the two overlapping bands belong to different sub-bands [see WS$_2$,  in Fig.\ref{fig:minibands} (a)] their corresponding Wannier functions will  differ in width (namely, the higher, more dispersive band, will have a larger spread). Thus in this case, the interaction parameters may differ significantly.  To account for this effect we consider the phenomenological parameter $\eta$ such that 
$U_{bb} = \eta U_{ab} = \eta^2 U_{aa}$. The interaction Eq.~\eqref{eq:H_I1} then assumes the form
\begin{align}\label{eq:H_I2}
\mc H _I = {U\over 2}\sum_{i}(\eta\,\delta n_{ia}+\delta n_{ib})^2\,,
\end{align}
$\eta<1$ describes the scenario where the Wannier function of band $b$ has a smaller spread when compared to $a$.  

The value of $U$ itself is twist-angle dependent~\cite{wu2018hubbard}.
For simplicity however, we will take a constant value $U=60$ meV for WS$_2$, WSe$_2$ and MoSe$_2$, which corresponds to a dielectric environment of $\ve = 5$~\cite{laturia2018dielectric}. For MoS$_2$ we  use $U=40$ meV. We note these values are  weaker than those used in other studies estimates~\cite{zhang2019moir,li2021imaging}.

The quadratic form of the interaction \eqref{eq:H_I2} was chosen for simplicity. In general, the ratio between the inter- and intra-species interactions is not controlled by a single parameter $\eta$. Therefore, it is important to note that the OSM phase space is expected to be reduced in the case where the interspecies interaction $U_{ab}$ becomes much larger than $U_{aa}$ or $U_{bb}$. As we  show in Appendix D and in the analysis of MoS$_2$ with =1however, the OSM state is not very sensitive to large interspecies interaction. Another crucial interaction we have neglected is longer range interaction. We expect these interactions to cause additional “Wigner crystal” insulating phases to appear. They will likely cause the phase space of the OSM state to shrink as well. However, these incompressible states may also stabilize over a finite range of doping with the aid of a background incompressible state, i.e. forming an orbitally selective wigner crystal.
Finally, we have also neglected spin exchange interactions (e.g. Hund's), which will be discussed in Appendix \ref{app:exc}.

\section{Details of the variational minimization of the slave-rotor mean-field free energy}\label{app:slave-rotor}
In this section we describe in more detail the slave-rotor mean-field theory~\cite{Florens2004slave,zaho2007self,chen2020competition}, that we have used in the main text.  
We first decompose the field operators 
into bosonic rotors ($e^{-i \t_{i\tau}}$) multiplied by neutral spinon operators ($f_{i\tau}$) 
\be 
\psi_{i\tau\s}=e^{-i\theta_{i\tau}}\,f_{i\tau\s} \,.
\ee
The respective density operator of each species, which are conjugates of the phases above, are then written in terms of angular momentum operators 
\begin{eqnarray}
\hat{L}_{i\tau}=-i\frac{\partial}{\partial\theta_{i\tau}}
\end{eqnarray}
subject to the {\it local} constraint
\[
\hat{L}_{i\tau}= f^\dagger_{i\tau}f_{i\tau}-1 \,,
\]
where the sum over spin is implicit. 
The application of the slave-rotor decomposition to Eq.\eqref{eq:H_b} and Eq.\eqref{eq:H_I} of the main text enables a simple mean-field analysis, which captures the localization-delocalization transition of a half-filled band.

This decomposition allows for a mean-field treatment of the Mott transition. The ``order parameter" is the quasi-particle weight $Z_{\tau}=|\langle e^{i\theta_\tau} \rangle|^2$. 
When the rotor's phase $\t_\tau$ assumes a finite expectation value, $Z_\tau\ne 0$ and the spinon quasi-particles have a finite overlap with the original electronic operator. This phase thus, corresponds to a Fermi-liquid state. On the other hand, when $\t$ is delocalized the rotor $e^{i\t_\tau}$ has a vanishing expectation value and the quasi-particle weight disappears ($Z_\tau=0$), corresponding to a Mott phase.  

Below we will describe two approaches. First, we will consider the case where only one species is decomposed (the flatter of the two). This case is more applicable to situation, where the density of states of the two bands differ significantly. In the second case, we will consider the same analysis where both bands are decomposed.

\subsection{Slave rotor decomposition of a single band in a two band system}
Applying the aforementioned slave-rotor decomposition to the flatter band 
(for the purpose of the discussion let it be, $\tau= b$), as is the case for Eq.\eqref{eq:H_b} and Eq.\eqref{eq:H_I} we obtain 
\begin{align}\label{eq:app:H_SR}
\mc H_{SR}=&- \sum_{ij}\left(t_b^{ij} e^{i(\theta_{ib}-\theta_{jb})}f^\dag_{ib}f_{jb}+t^{ij}_a \psi^\dag_{ia}\psi_{ja}\right)
\\
+&\sum_i \left(\bar \e_b f^\dag_{ib} f_{ib}+\bar\e_b \psi^\dag_{ia} \psi_{ia}\right)+{U\over 2}\sum_{i} \left( \hat L_{ib} + \eta \delta n_{ia} \right)^2
\nn
\end{align}
where $t^{ij}_\tau$ are the set of tight-binding parameters that reproduce the dispersive part $\e_k$ in Eq.\eqref{eq:disp} when transformed to reciprocal space.

To estimate the location of possible Mott phases of Eq.\eqref{eq:app:H_SR} we use a variational approach. This should be cotrasted with Ref.~\onlinecite{Florens2004slave} where the self-consistent mean-field technique was used.  In the variational approach we minimize the expectation value of Eq.~\eqref{eq:app:H_SR} with respect to a variational wave function denoted by $|\W_V\rangle = |K_b,h_b\rangle\otimes|\mu_b\rangle\otimes |\mu_a\rangle$, which is a product of the ground-states of the following variational Hamiltonians  
\begin{align}
&H_{\theta}^b=\frac{1}{2}\hat{L}^2_{b}+h_b\hat{L}_{b}+K_{b}\cos\theta_{b}\label{eq:app:H_theta}\\
&H_{f}^b=\sum_{k}\left(\e_{kb}-\mu_{b}\right)f^\dagger_{kb}f_{kb}\label{eq:H_f}\\
& H_\psi^a = \sum_k\left(\e_{ka}+\D \bar \e-\mu_{a}\right)\psi^\dagger_{ka}\psi_{ka}\,,\label{eq:H_psi}
\end{align}
where we have shifted the energies such that the center of band $b$ is at zero and $\D \bar \e = \bar \e_a - \bar \e _b$.
Eq.\eqref{eq:app:H_theta} controls the rotor field, where the term proportional to $K_b$ acts to pin the phase $\t_b$ giving rise to a finite quasi-particle weight $Z_b$. Thus, we can identify the Fermi-liquid (Mott) phases with situations where the minimal energy solution is obtained with $K_b \ne 0$ ($K_b = 0$). The parameter $h_b$ is used to obey the slave-rotor constraint on average.
The second and third variational Hamiltonians Eq.~\eqref{eq:H_f} and \eqref{eq:H_psi}  generate  Fermi sea states of spinons and $b$-electrons, with density controlled by the parameter $\mu_a$ and $\mu_b$, respectively. Notice that ground state of Eq.~\eqref{eq:H_f} is independent of the band width and therefore $Z_b$ is omitted.

We then minimize the expectation value of the full Hamiltonian Eq.\eqref{eq:H_SR}, denoted by
\[
F(K_b,h_b,\mu_b,\mu_a) = \langle \W_V|\mc H_{SR}|\W_V\rangle\,,
\]
with respect to the four parameters $K_b$, $h_b$ and $\mu_b$ and $\mu_a$ subject to two constraints 
\be\label{eq:con}
\langle \hat L_b \rangle = \langle f^\dag_{ib}f_{ib} \rangle-1
\;\;; \;\;
\langle  f^\dag_{ib} f_{ib}\rangle+\langle\psi^\dag_{ia}\psi_{ia} \rangle = n
\ee
The difference between the number of constraints and variational parameters implies that two are free. These correspond to the  the quasi-particle weight of band $b$ and any distribution of the total density between the bands. These two parameters are dictated by energetics.

Notice that in using Eq.\eqref{eq:H_theta} we have neglected spatial fluctuations of the field  $\t_b$. This restricts our ground state manifold (for example, it can not capture spin-correlations~\cite{zaho2007self}). However, it allows for a  significant simplification: The expectation value of the rotor correlation function becomes a product of local expectation values $\langle e^{i(\t_{i\tau}-\t_{j\tau})}\rangle = \langle e^{i\t_{i\tau }}\rangle\langle e^{-i\t_{j\tau}}\rangle=Z_\tau$. Consequently, the expectation value of the kinetic energy terms can be straightforwardly transformed back to momentum space, reproducing the exact continuum dispersion relation Eq.\eqref{eq:H_b}.
\begin{align}
    F = &\sum_{k}\left[ Z_b \epsilon_{kb}N_{kb} + (\epsilon_{ka}+\D\bar \e)N_{ka} \right] \label{eq:F}
    \\
    &+{U\over 2}\sum_i \left[ \langle {L_b^2}\rangle+2\eta \langle L_b \rangle \langle  n_a - 1\rangle +\eta^2  \langle ( n_a - 1)^2\rangle  \right]\,,\nn
\end{align}
 where $N_{k b} = N_0(\e_{kb}-\mu_b)$, $N_{ka} = N_0(\e_{ka}+\D\bar \e-\mu_a)$ and $N_0(x) = 1/(e^{\b x}+1)$. Here $\b$ is the inverse temperature, which will be taken to infinity $\b \to \infty$, which is used as a numerical parameter to smoothen the discretization.  

\begin{figure}[!b]
  \centering
  \includegraphics[width=0.5\textwidth]{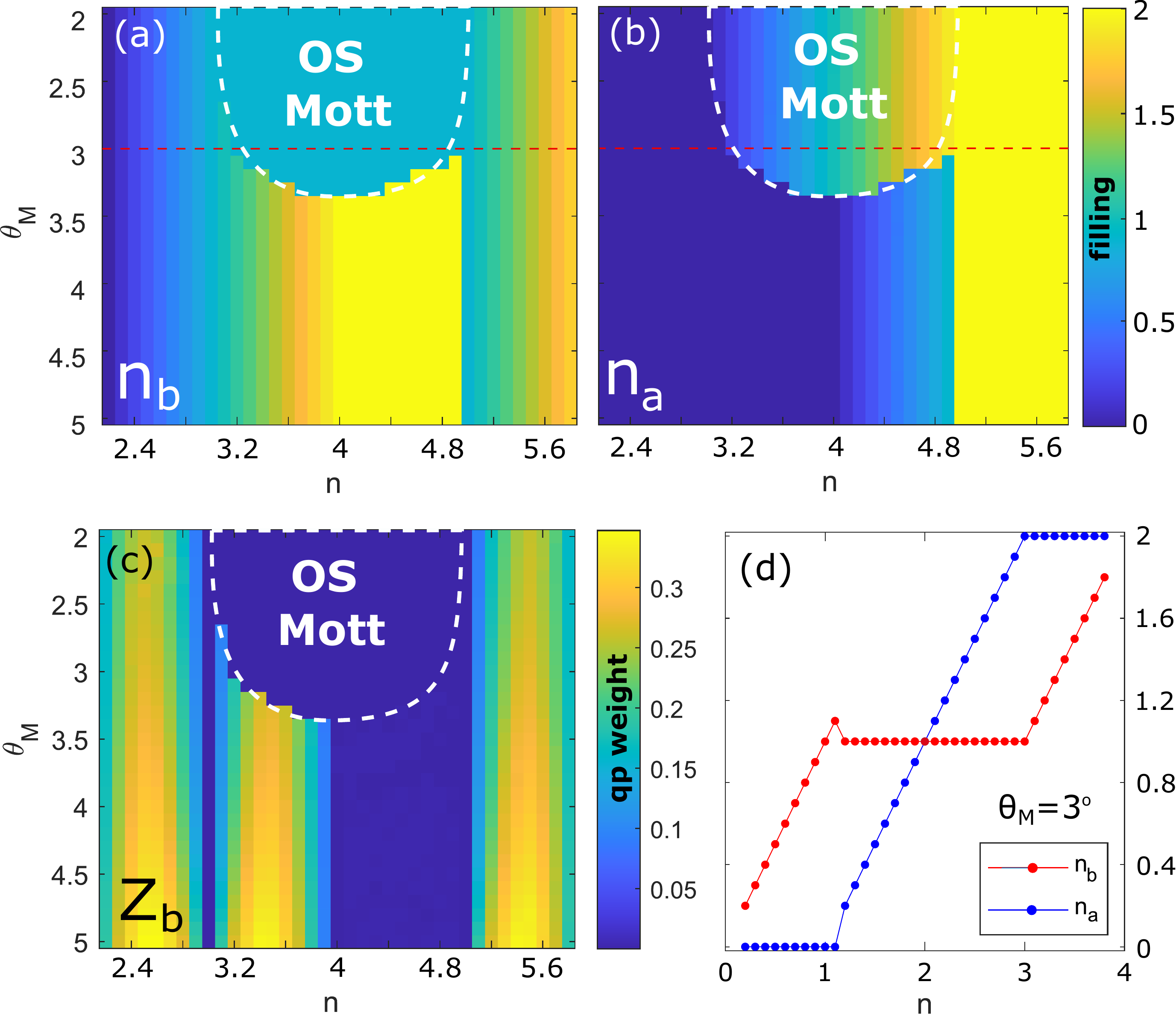}
  \caption{The results of the slave rotor mean-field analysis for MoSe$_2$, $U = 60$ meV and $\eta = 1/2$. (a) and (b) The density of the two species in the space of the moir\'e angle $\t_M$ and total density $n$. (c) The quasiparticle weight of species $b$, $Z_b$. The region of half-filling $n_b = 1$ and $Z_b = 0$ corresponds to a Mott state of band $b$. (d) The densities of bands $a$ and $b$ at $\t_M$ vs. total density. \label{fig:app:PDMoSe2}}
\end{figure}

\begin{figure}[!b]
  \centering
  \includegraphics[width=0.5\textwidth]{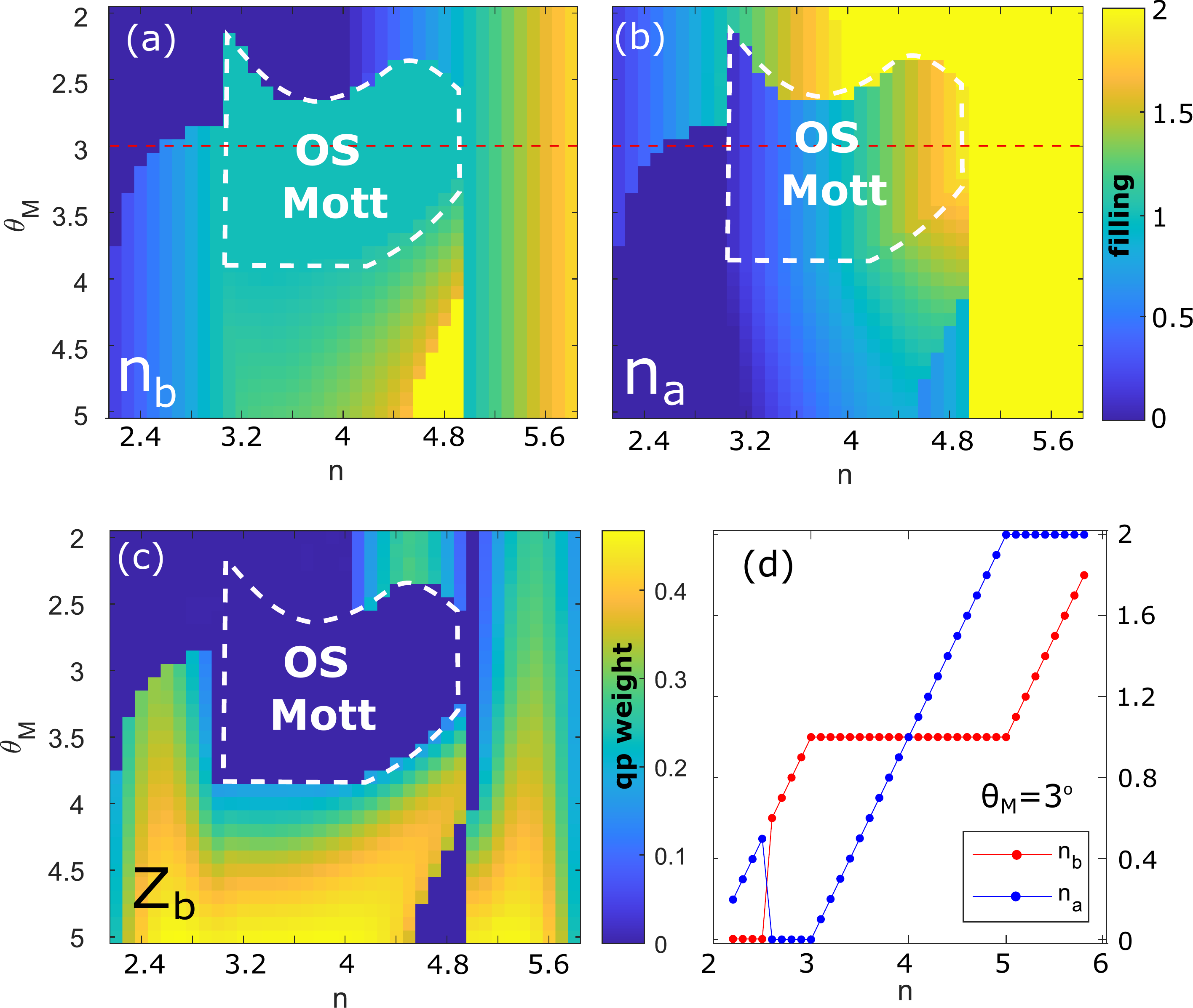}
  \caption{The results of the slave rotor mean-field analysis for WSe$_2$, $U = 60$ meV and $\eta = 1/2$. (a) and (b) The density of the two species in the space of the moir\'e angle $\t_M$ and total density $n$. (c) The quasiparticle weight of species $b$, $Z_b$. The region of half-filling $n_b = 1$ and $Z_b = 0$ corresponds to a Mott state of band $b$. (d) The densities of bands $a$ and $b$ at $\t_M$ vs. total density. \label{fig:app:PDWSe2}}
\end{figure}

In panels (a)-(d) of Fig.\ref{fig:phase_diagram}  and Figs.~\ref{fig:app:PDMoSe2}, \ref{fig:app:PDWSe2} we plot 
the phase diagram obtained from minimizing Eq.~\eqref{eq:F} for the band structure parameters of WS$_2$, MoSe$_2$ and WSe$_2$, respectively. 
Note that as opposed to the main text we do not specify the precise bilayer composition. For each TMD material here, one must consider a second ``inactive" layer that induces the moir\'e potential and has band alignment properties that ensure it has a higher-in-energy conduction band. 
We use $U = 60$ meV and $\eta = 1/2$. Panel (a) and (b) are maps of the density of bands $b$ and $a$, respectively, in the space of total density $n$ and twist angle $\t_M$. Panel (c) is the corresponding quasi-particle weight $Z_b$. Panel (d) shows the relative filling at $\t_M = 3$. 
There are two distinct regimes as a function of angle. 
For $\t_M < 3.5$ the filling of band $b$ is roughly split in half. Between $n=0$ and $n = 1$ band $b$ fills until it reaches a localized state (characterized by $Z = 0$ and $n_b=1)$. Then band $b$ fills continuously between $n=1$ to $3$. Finally, band $b$ continues to fill until $n=4$ is reached.  The regime where band $a$ is continuously filling realizes an orbitally selective Mott phase, where a Kondo lattice model is expected to be simulated with variable itinerant electron density.

On the other hand, for $\t_M>3.5$ the bands fill up one-by-one. In particular, band $b$ fills completely between $n=2$ and $n=4$ with a Mott state at $n=5$. Then above $n = 5$ it resets back into the Mott state and band $a$ fills completely. This behavior thus, resembles a Stoner-like polarization of the species. However, we comment that the slave-rotor mean field tends to overestimate the size of band-polarized regions. This is because it overestimate the contribution of charge fluctuations to the interaction energy when the filing differs significantly from $1/2$.

\subsection{Two band slave rotor decomposition}
As explained in the case of MoS$_2$ the electronic bands experience a much weaker spin-orbit coupling. Consequently, the shape and effective binding energies are almost identical [see Table~\ref{tab:table1} and Fig.\ref{fig:minibands} (b)]. 
In this case it makes sense to decompose both bands in an unbiased manner
\begin{align}\label{eq:H_SR2}
\mc H_{SR}=&-\sum_{ij\tau}t_\tau^{ij} e^{i(\theta_{i\tau}-\theta_{j\tau})}f^\dag_{i\tau}f_{j\tau}\\&-\sum_{i\tau}
\bar \e_\tau f^\dag_{i\tau} f_{i\tau}  +\frac{U}{2}\sum_{i}\left(L_{ia}+L_{ib} \right)^2
\nn
\end{align}
where $\eta = 1$ and $\D\bar \e = 
\bar \e_a-\bar \e_b$ is much smaller than the band width and $U$. 

Following the previous section, we now use four variational Hamiltonians of the form Eq.\eqref{eq:H_theta} and \eqref{eq:H_f}, which generate a ground state $|K_\tau,h_\tau,\mu_\tau\rangle$ controlled by six variational parameters and subject to three constraints 
\be\label{eq:con2}
\langle \hat L_\tau \rangle = \langle f^\dag_{i\tau}f_{i\tau} \rangle-1
\;\;; \;\;
\langle  f^\dag_{ia} f_{ia}\rangle+\langle\f^\dag_{ib}\f_{ib} \rangle = n
\ee

In panels (e)-(h) of Fig.\ref{fig:phase_diagram} of the main text we plot the phase diagram resulting from the minimization of the expectation value of Eq.~\eqref{eq:H_SR2}. As can be seen at $n = 2$ both $Z_a$ and $Z_b$ equal zero for small enough angles. The similarity of the bands of the two species leads us to propose electron doped MoS$_2$ as a candidate material to realize an SU(4) symmetric Mott insulator on a triangular lattice, which is an interesting problem on its own right which can is an interesting situation on its own right~\cite{keselman2020dimer,keselman2020emergent}.   

\subsection{Details of the numerical minimization procedure}
In this appendix we provide the details of the numerical minimization procedure of Eq.~\eqref{eq:F}. 
To calculate this functional, we performed straightforward Brillouin zone integration on a square grid of size 150$\times$150.
The integration itself was performed by matlab's {\it trapezoidal} numerical integration, and the fermi-dirac distribution was written as:
$N_{0}(\epsilon_k)=\frac{1}{1+e^{-\beta\epsilon_k}}$
With $\beta=\frac{1}{T}$ being the inverse temperature. To broaden the discretization we use a finite temperature $\beta=60/max(\epsilon_a)$.
In addition, the minimal ground state that was found for the phase diagram in Fig.3 was found by matlab's minimization algorithm {\it fmincon}, which minimizes the functional  Eq. ~\eqref{eq:F}, under the constrains Eq. ~\eqref{eq:con} by means of the specified variational parameters.
The optimization algorithm that was found to converge most efficiently was the {\it interior-point}  algorithm, which is the default algorithm of fmincon.

\section{On the stability of the OSM state to variations in of phenomenological parameters}\label{app:stab}
In the main text we have presented the results of a slave-rotor mean-field analysis, where bands of different species fill either one by one or simultaneously, depending on the twist angle. 
The latter scenario is is of particular interest to us as it gives way to the orbitally selective Mott phase. 

Given that we have a number of unknown parameters, including $\eta$ and the energy difference 
\be 
\D \bar \e = \bar \e_a - \bar \e_b\,,
\ee
it is important to test the stability of the OSM state. 
In this appendix we compute a lower bound on the phase space volume of the OSM state in the space of $\D \bar \e$ and $\eta$. 

To obtain this estimate we focus specifically on the commensurate  filling $n = 2$ (or $n=4$ for the strongly spin-orbit coupled bilaer WS$_2$/WSe$_2$). If  the bands fill one-by-one this filling point will be characterized by one completely filled band and another completely empty. On the other hand, if the bands fill simultaneously this filling value is likely to be characterized by a partial filling of both bands with total density of unity. Here we will make a restrictive assumption, that both bands are at filling unity, where one is in a Mott state and the other is either metallic or also in a Mott state. The phase diagrams we have computed are consistent with this behavior at  filling $n=2$ (or 4). 

Under this assumption we can estimate the expectation value of the Hamiltonian Eq.\ref{eq:H_SR} within these restrictive trial states:
\begin{align}
&|F_a\rangle = |n_a = 2\rangle \otimes |n_b=0\rangle \\
&|F_b\rangle = |n_a = 0\rangle \otimes |n_b=2\rangle\nn \\
&|P\rangle = |n_a = 1\rangle \otimes |n_b=1\rangle\nn
\end{align}
and compare which of them has a lower energy. 
The first two states represent fully polarized states and thus corresponds to the scenario where the band fill one-by-one. The third state however, is where the filling is shared between the two bands. We will further assume that band $a$ is the flatter of the two bands, and is in a Mott state at $n_a = 1$. 

The energy per unit cell of these three states is given by \begin{align}\label{eq:app:Es}
&E_a \equiv {\langle F_a | \mc H_{SR}|F_a \rangle\over \W}  = 2\bar \e_a+{U\over 2}(1-\eta)^2\\
&E_b \equiv {\langle F_b | \mc H_{SR}|F_b \rangle\over \W}  = 2\bar \e_b+{U\over 2}(1-\eta)^2\nn \\
&E_{ab}\equiv {\langle P | \mc H_{SR}|P \rangle\over \W} =  \bar \e_a+\bar \e_b+ T_{it}\nn
\end{align}
Here $\W$ is the total number of sites. $T_{it}\leq 0$ is the sum of the negative kinetic energy  associated with half-filling band $b$, and the interaction energy associated with the charge fluctuations $\D n_b^2=\langle (n_b-1)^2\rangle$ at the half-filling point. Thus, when $T_{it}<0$ band $b$ remains in a metallic state and reaches zero when it falls into a Mott state as well. 

When $E_{ab}<\mrm{min}(E_a,E_b)$ the third state (simultaneous filling) is more favourable energetically.
On the other hand, when $E_a$ or $E_b$ are minimal, the ground state {\it can } be band-polarized (but not necessarily), where the bands fill one-by-one. 

\begin{figure}[!b]
  \centering
  \includegraphics[width=0.4\textwidth]{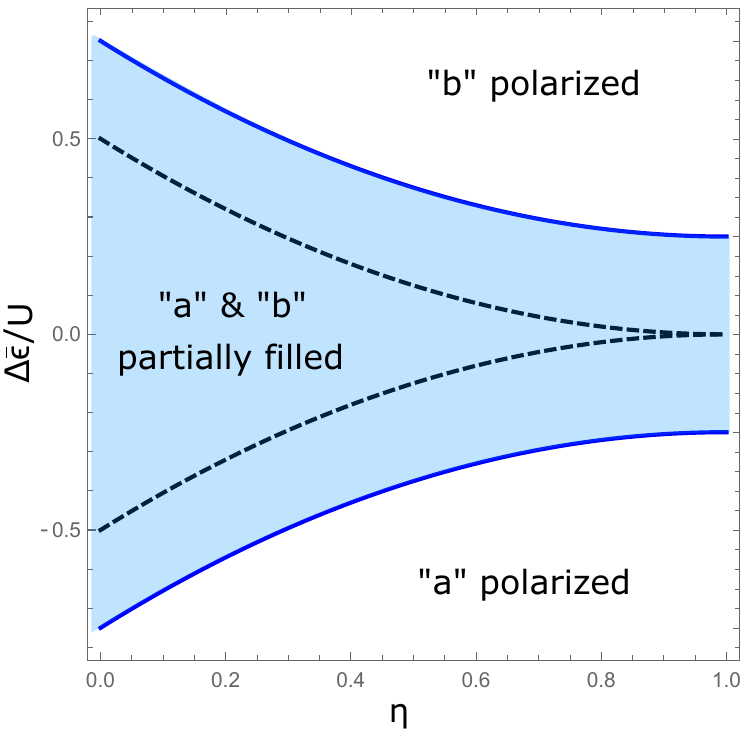}
  \caption{Stability of simultaneous occupation of two moir\'e minibands for filing $n = 2$ (or $n=4$ for the strongly spin-orbit coupled bilayer WS$_2$/WSe$_2$) in the space of $\D \bar \e$ and the phenomenological parameter $\eta$. The blue lines correspond to $T_{it} = -U/4$ (species $b$ in a metallic state) and the dashed lines are the case $T_{it} = 0$ (species $b$ also in a Mott state).   \label{fig:app:stab}}
\end{figure}

Comparing these energies, we conclude that the partial filling state is stable at least in the regime 
\be\label{eq:app:stab}
|\D \bar \e| < U(1-\eta)^2/2-T_{it} \,.
\ee
In Fig.~\ref{fig:app:stab} we plot the stability region defined by Eq.~\eqref{eq:app:stab} for the case of $T_{it} = -U/4$. The case of $T_{it} = 0$ is also plotted for comparison (dashed line).

Given that $T_{it} \leq 0$, there exists such a regime for any value of $\eta$. Note that the width in $\D \bar \e$ of this window scales with $U$ at strong coupling. We thus conclude that the regime of partial occupation of both bands is wide and robust to parameters such as the ratio of species interaction $\eta$ and splitting of the bands $\D \bar \e$.

\begin{figure}
  \centering
  \includegraphics[width=0.45\textwidth]{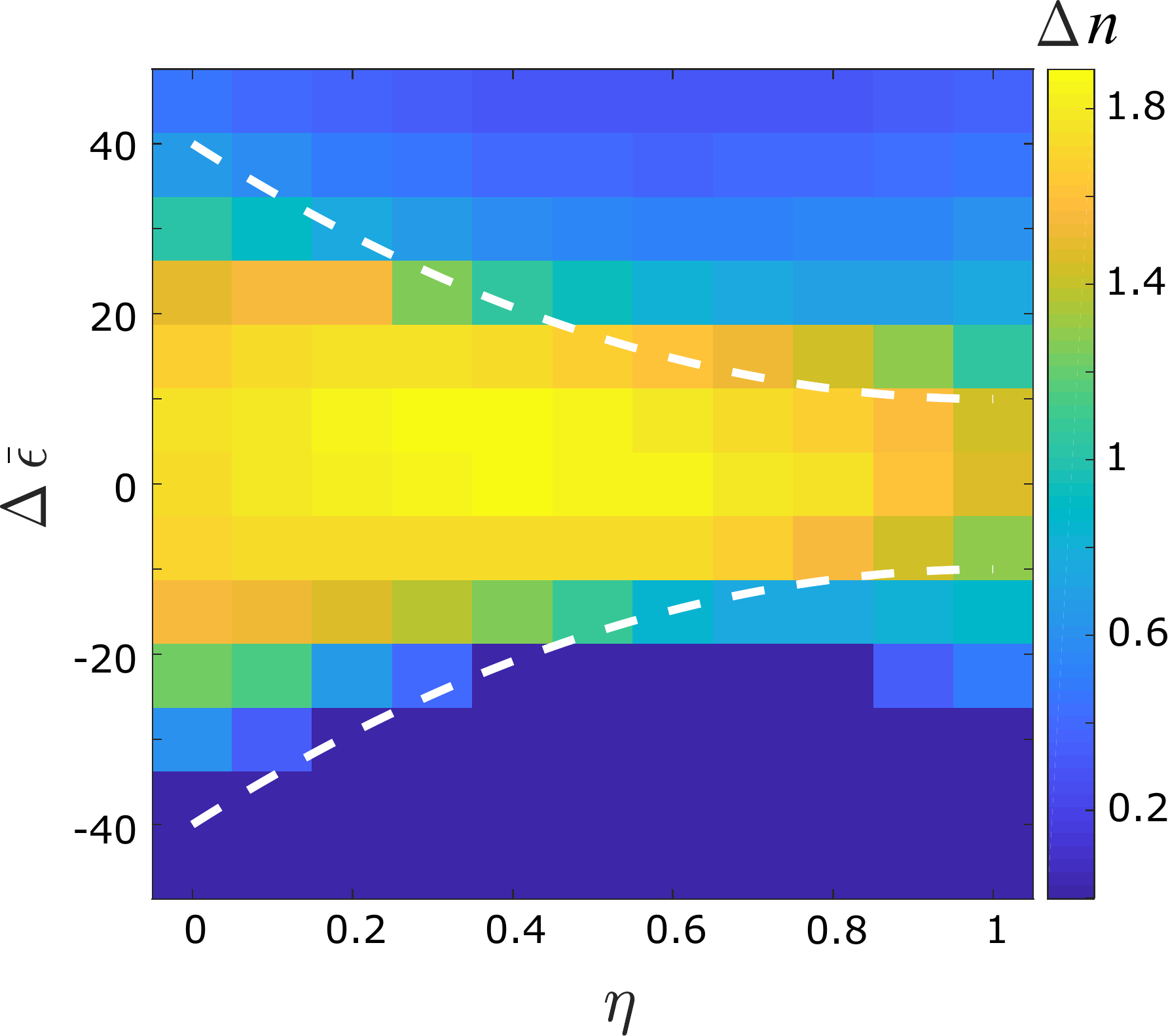}
  \caption{ Stability diagram of the coexistence of localized and itinerant states in the space of $\eta$ and $\D \bar \e$ with $U = 60$ meV obtained from numercial minimization of Eq.~\eqref{eq:F}. The colorbar denotes the width of the Mott region in filling averaged over all angles between $2^\circ$ and $5^\circ$ (the maximal width is 2). The  white dashed line is the analytic estimate, where the two species are expected to be simultaneously at partial filling at $n = 4$ Eq.~\eqref{eq:app:stab} with $T_b = -10$ meV.     \label{fig:stab:main}}
\end{figure}

In Fig.~\ref{fig:stab:main} we plot the width of the OSM phase $\D n = n_{max}-n_{min}$ averaged over the angles $2^\circ$ and $5^\circ$, which is obtained from the numerical minimization of Eq.~\eqref{eq:F}. 
Here $n_{max}$ and $n_{min}$ mark the boundaries of the OSM phase per angle (maximal value is 2), as shown in Fig. \ref{fig:phase_diagram} .(c). 
The white dashed line is the analytic estimate Eq.~\eqref{eq:app:stab}. 
We thus, conclude that the OSM phase is not sensitive to parameters and is a generic feature of the phase diagram of electron doped moir\'e TMDs.

\section{Spin Exchange interactions and expected phenomenology}\label{app:exc}
Inside the OSM phase charge fluctuations of the localized species are quenched and therefore, the relevant inter-species interactions are of spin-exchnage type.  
In this section we discuss two such interactions and their possible influence on the magnetic ground state.

We anticipate that the ferromagnetic Hund's coupling is the largest exchnage mechanism 
\be \label{eq:Hunds}
\mc H_H = -\sum_i J_H (\psi_{ib}^\dag \bs \s \psi_{ib})\cdot \bs S_i
\ee
where $\bs S_i$ is the spin of the localized moments at site $i$ (let us assume they belong to species $a$) . Using standard harmonic oscillator states, we estimate $J_H \approx 0.2 U$. 

Upon approaching the meting point of the OSM state however, other interaction terms that compete with Eq.~\eqref{eq:Hunds}. For example, the interaction that scatters across the original Brillouin zone. To see this let us first write this term in basis of the original operators  Eq.~\eqref{eq:H0}
$
\mc H_J =J_{P}\sum_{k,k',p}c^\dag _{k+p\,a\uparrow}c^\dag _{k'-p\,a\downarrow}c _{k'\,b\downarrow}c _{k\,b\uparrow}+h.c.
$, 
where $J_{P}/U \sim a/a_M \approx \sqrt{\d^2 +\t_M^2 }$ (for an angle of $\t_M = 3.5^\circ$ we obtain $J_P \approx 0.1 U$). 
Taking into account the moir\'e potential and the slave-rotor decomposition described above, this  interaction assumes the form
\be\label{eq:HJ2}
\mc H_J =\tilde J_P\sum_{k,k',p} f^\dag _{k+p\,a\uparrow}f^\dag _{k'-p\,a\downarrow}\psi _{k'\,b\downarrow}\psi _{k\,b\uparrow}+h.c.
\ee
where $\tilde J_P = Z_a J$.
Thus, when the band $a$ is localized and $Z_a=0$ this term vanishes. However, if the local moments are incorporated into the Fermi surface (through the formation of a heavy Fermi liquid) they reacquire a finite quasi-particle weight $Z$~\cite{chen2020competition}. Thus, this interaction can become important close to the melting point of the local moment lattice. 

As mentioned above, inside the OSM phase  we expect the dominant exhnage to be Hind's and therefore spin-correlations to be  ferromagnetic as in Ref.~\onlinecite{anisimov2002orbital}. 
When the Hund's coupling Eq.~\eqref{eq:Hunds} is dominant the main influence of the itinerant electrons inside the OSM state is to mediate long-ranged RKKY interactions~\cite{fischer1975magnetic} 
\be\label{eq:RKKY}
\mc H_{\rm{RKKY}} \approx {J_H^2 \nu_0 k_F^2 \over 8\pi} \sum_{ij}{\sin k_F R_{ij}\over R_{ij}^2}\bs S_i \cdot \bs S_j
\ee
where $\nu_0$ is the density of states. 
Thus, as the filling of the itinerant band is modified from zero to 2, the {\it nearest-neighbour} interaction mediated by the electrons can be tuned from ferromagnetic in the dilute limit to antiferromagnetic and back to ferromagnetic (going through a van Hove singularity). This interaction is added to the direct superxchnage between sites~\cite{wu2018hubbard}, which may have a cooperative effect or frustrate the magnetic interactions.

\section{\label{sec:Kondo} The heavy Fermi liquid state }
When the antiferromagnetic correlations dominate we expect a heavy-Fermi liquid state to compete with the internal magnetic interactions. For completeness, in this appendix we compute the Kondo temperature assuming $\tilde J_P = 4$ meV within the large-N mean-field theory~\cite{hewson1997kondo}. We find hat $T_K\sim$5-15 K inside the OSM phase and therefore, we expect that if the antiferromagnetic correlations dominate the OSM melting transition can be accompanied by a detectable heavy Fermi liquid state. Moreover, in this case the tunability of the itinerant electron density may allow to tune though the Doniach phase diagram~\cite{doniach1977kondo}.

\begin{figure}
  \centering
  \includegraphics[width=0.25\textwidth]{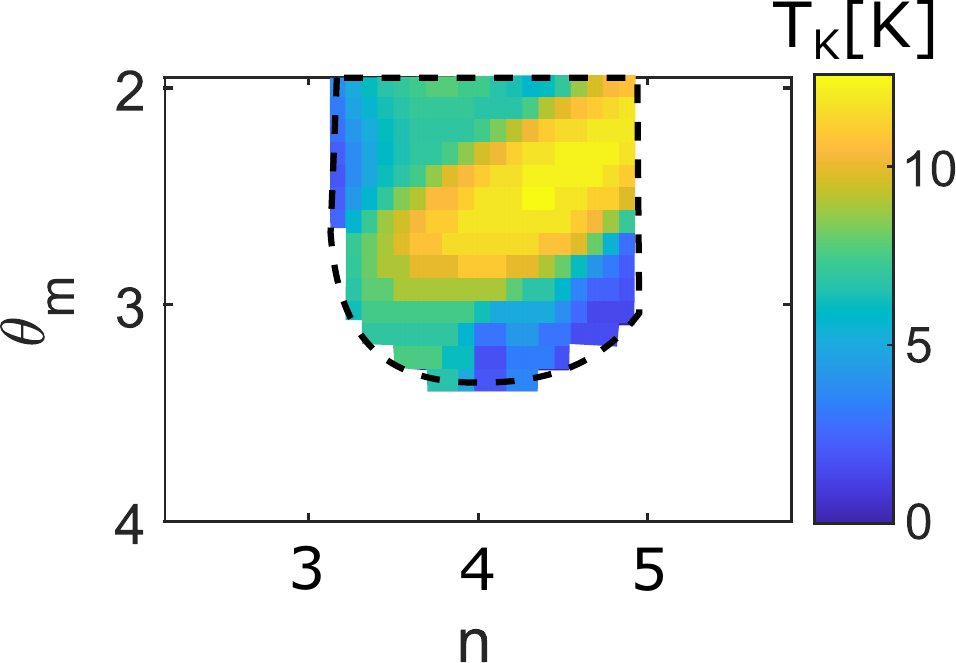}
  \caption{  The Kondo temperature inside the OSM state for MoSe$_2$.  \label{fig:kondo}}
\end{figure}

Let us breifly describe the large-N mean field theory. The dispersion of the two species is taken to be
\be
\mathcal{H}_{MF} = \sum_{ k}\left[  (Z_b\e_{kb}-\mu_b)f_{k b}^\dag f_{kb}+(\e_{ka}-\D\bar \e-\mu_a)\psi_{ka}^\dag \psi_{ka}\right]
\ee
where the density of each species $n_\tau$ is set separately using the Lagrange multipliers $\mu_\tau$ according to their values in the slave-rotor mean-field calculation. 
We then decouple Eq.~\eqref{eq:HJ2} using the mean-field hybridization $\chi ={J/2} \langle f_{b\s}^\dag\psi_{a\s}\rangle + c.c.$.  
We then solve for $\chi$ self-consistently, while tuning $\mu_a$ and $\mu_b$ to conserve the density of $a$ and $b$ on average. The Kondo temperature is then estimated by seeking the lowest temperature where the self-consistent solution for $\chi = 0$. To obtain the Kondo temperature $T_K$ we estimate the lowest temperature where $\chi = 0$. 

In Fig.~\ref{fig:kondo} we plot the resulting Kondo temperature as a function of twist-angle and density for $\tilde J = 4$ meV. As can be seen, the Kondo temperature measurable in standard cryosthetics and might be physically important.

%

\end{document}